# Acoustic corner states in topological insulators with built-in Zeeman-like fields

Xueqin Huang[1*], Jiuyang Lu[1*], Zhongbo Yan[2], Mou Yan[1], Weiyin Deng[1†], Gang Chen[3,4†], and Zhengyou Liu[5,6†]

[1]School of Physics and Optoelectronics, South China University of Technology, Guangzhou 510640, China
[2]School of Physics, Sun Yat-Sen University, Guangzhou 510275, China
[3]State Key Laboratory of Quantum Optics and Quantum Optics Devices, Institute of Laser spectroscopy, Shanxi University, Taiyuan 030006, China
[4]Collaborative Innovation Center of Light Manipulations and Applications, Shandong Normal University, Jinan 250358, China
[5]Key Laboratory of Artificial Micro- and Nanostructures of Ministry of Education and School of Physics and Technology, Wuhan University, Wuhan 430072, China
[6]Institute for Advanced Studies, Wuhan University, Wuhan 430072, China

*X.H. and J.L. contributed equally to this work
†Corresponding author. Email: dengwy@scut.edu.cn; chengang971@163.com; zyliu@whu.edu.cn

**The higher-order topological insulators (HOTIs), with such as the topological corner states, emerge as a thriving topic in the field of topological physics. But few connections have been found for the HOTIs with the well explored first-order topological insulators described by the $\mathbb{Z}_2$ index. However, most recently, a proposal asserts that a significant bridge can be established between the HOTIs and the $\mathbb{Z}_2$ topological insulators. When subject to an in-plane Zeeman field, the corner states, the signature of the HOTIs, can be induced in a $\mathbb{Z}_2$ topological insulator. Such Zeeman field can be produced, for example, by the ferromagnetic proximity effect or magnetic atom doping, which obviously involves the drastically experimental complexity. Here we show that, a phononic crystal, designed as a bilayer of coupled acoustic cavities, hosts exactly the Kane-Mele model with built-in in-plane Zeeman fields. We observe that the helical edge states along the zigzag edges are gapped, and the corner states, localized spatially at the corners of the samples, appear in the gap, confirming the effect induced by the Zeeman field. We further demonstrate the intriguing contrast properties of the corner states at the outer and inner corners in a hexagonal ring-shaped sample.**

Two-dimensional (2D) $\mathbb{Z}_2$ topological insulators (TIs), hosting a pair of gapless helical or spin-filtered edge states on the boundary protected by time-reversal symmetry[1,2], have attracted wide attention in electronic, photonic, and phononic materials. The $\mathbb{Z}_2$ TIs are first demonstrated in the Kane-Mele (KM) model[3] and the Bernevig-Hughes-Zhang model[4], and lately predicted to exist in a large fraction of nonmagnetic materials[5-7]. The KM model is a hexagonal honeycomb lattice endowed with intrinsic spin-orbit coupling. Although the KM model is difficult to be implemented in practice due to the extremely weak strength of the intrinsic spin-orbit coupling[8,9], it captures all the essentials of the $\mathbb{Z}_2$ TI, and is ideal for exploring the associated topological properties and phase transitions.

The HOTIs generalize the concept of TIs with unconventional bulk-boundary correspondence[10-13]. The 2D second-order TIs are typical HOTIs, which are characterized by the 0D in-gap corner states[13], rather than the 1D gapless edge states. The 2D second-order TIs have more than one topological origin. The first is attributed to the nested Wilson loops or the related quantized quadrupole moments, thus is also called quadrupole TI[13-18]. The second is the Wannier-type second-order TI, whose topological invariant is the quantized Wannier centers[19-25]. Searching for the new mechanisms[26-31], such as the Jackiw-Rebbi mechanism[30-32], for the second-order TIs still remain a hot topic. Very recently, a significant advance has been achieved, and the corner states were predicted in a conventional 2D $\mathbb{Z}_2$ TI subject to the applying in-plane Zeeman fields[33,34], which builds a bridge between the $\mathbb{Z}_2$ TIs and second-order TIs. However, the in-plane Zeeman field, possibly produced by ferromagnetic proximity effect or magnetic atom doping[34-36], drastically increases the experimental complexity. It is a great challenge to observe such corner states experimentally in the $\mathbb{Z}_2$ TIs.

Phononic crystal (PC) offers an ideal platform for investigating topological physics[37-40], because of its designable macroscopic structure, which allows the direct observation of such corner states. We construct a bilayer PC consisting of coupled acoustic cavities in hexagonal lattice. The synthetic intrinsic spin-orbit coupling and the Zeeman-like field are both induced by the chiral interlayer couplings, with the layer degree of freedom as pseudospin. As such, the Zeeman-like field is intrinsic to the PC, with the direction along the $y$ axis. The gapped edge states and in-gap topological corner states are observed evidently in a hexagon-shaped PC, revealing itself as a second-order TI. The unique properties of the corner states at the outer and inner corners of a hexagonal ring-shaped PC sample are demonstrated.

We first introduce a tight-binding model of a bilayer hexagonal lattice, which is a modified KM model, i.e., the KM model with a Zeeman field along the $y$ direction. Figure 1a shows the hexagon-shaped sample of the lattice model, where a primitive cell contains two layers with two nonequivalent sites in each layer, denoted by A (red) and B (blue) as shown in Fig. 1b. The nearest neighbor hoppings of the intralayer (gray) are $t_0$, and the chiral hoppings of the interlayer (cyan) are $t_c$. On the basis of $(A_\uparrow, B_\uparrow, A_\downarrow, B_\downarrow)$, where $\uparrow (\downarrow)$ represents the pseudospin of upper (lower) layer, the Bloch Hamiltonian in momentum space can be written as

$$H_0 = \begin{pmatrix} 0 & t_0 f & t_c g & 0 \\ t_0 f^* & 0 & 0 & t_c g^* \\ t_c g^* & 0 & 0 & t_0 f \\ 0 & t_c g & t_0 f^* & 0 \end{pmatrix}, \tag{1}$$

where

$$f = 1 + 2e^{-i\sqrt{3}\,k_y/2} \cos k_x/2,$$

$$g = e^{-ik_x} + 2e^{ik_x/2} \cos \sqrt{3} k_y/2,$$

and the lattice constant is set to the unit length. The chiral interlayer couplings open a bulk gap, as shown in Fig. 1c. The low energy effective Hamiltonian around the $K$ and $K'$ points is

$$H_k = v_\mathrm{F}(k_x \sigma_x \tau_z + k_y \sigma_y) + V_\mathrm{SO} \sigma_z s_y \tau_z + V_\mathrm{Z} s_x, \tag{2}$$

where $v_\mathrm{F} = -\sqrt{3} t_0/2$, $V_\mathrm{SO} = -3\sqrt{3} t_c/2$, $V_\mathrm{Z} = -3 t_c/2$, and $\boldsymbol{\sigma}, \boldsymbol{s}, \boldsymbol{\tau}$ represent the sublattice, layer and valley degrees of freedom. To compare with the KM model, we make a unitary transformation $H_u = U H_k U^\dagger$ with

$$U = \frac{1}{\sqrt{2}} \begin{pmatrix} 1 & -i \\ 1 & i \end{pmatrix},$$

where the matrix is spanned in layer degree of freedom, and obtain the Hamiltonian

$$H_u = v_\text{F}(k_x\sigma_x\tau_z + k_y\sigma_y) + V_\text{SO}\sigma_z s_z\tau_z + V_\text{Z} s_y. \quad (3)$$

One can see that the first two terms are the KM Hamiltonian with the intrinsic spin-orbit coupling[3], and the third term is the Zeeman field along the $y$ direction. So this bilayer hexagonal lattice is equivalent to the modified KM model, in which the chiral interlayer couplings provide both the effective intrinsic spin-orbit interactions and the built-in Zeeman-like field along the $y$ direction.

The corner states are predicted to be induced by the in-plane Zeeman field in the $\mathbb{Z}_2$ TIs[33,34], thus should exist in the bilayer hexagonal lattice. We first calculate the projected dispersion of a ribbon with the zigzag boundaries, as shown in Fig. 1d. The synthetic Zeeman field breaks the spin-1/2 time-reversal symmetry, and results in the gapped edge states with an effective Hamiltonian $H_e = vqs_z + ms_y$, where $v$ is the velocity term, $q$ is the momentum long the edge direction, and $m$ is the mass term induced by the synthetic Zeeman field[34]. There are two kinds of configurations of zigzag boundary, corresponding to the outermost site being A (red) and B (blue) respectively, as shown by the dashed box in Fig. 1a. Although the edge dispersions of these two zigzag boundaries are same, the edge modes at $k_x = \pi$ are inversed with respect to the $c_{2y}$ symmetry (two-fold rotation about the $y$ axis), indicating the edge Hamiltonians host the opposite mass signs, i.e., if the edge state along one boundary has a positive mass $m$, the other one is $-m$.

According to the Jackiw-Rebbi mechanism, the corner states, as the domain wall modes, emerge between the two topologically distinct zigzag boundaries with opposite masses. Figure 1e shows the eigenvalues for the hexagon-shaped sample (Fig. 1a), in which the distinct zigzag boundaries are intersected at the corners. The existence of six corner modes inside the gap of edge states corresponds with the above topological discussion and indicates this system is a second-order TI. Interestingly, the corner states mainly locate at the right of the corner for the upper layer, but at the left for the lower layer, as shown by the inset in Fig. 1e. We further calculate the eigenvalues of the hexagon-shaped sample as a function of $t_c/t_0$, as shown in Fig. 1f. The system hosts second-order TI phase with corner states (red lines) for $0 < t_c/t_0 < 1$, and turns to the semimetal phase for $t_c/t_0 \geq 1$ (Fig. S1 in Supplementary Information). It is worth noting that the corner states remain robust in the absence of the two-fold rotation symmetry, as long as the gaps of bulk and edge states stay open, where the Jackiw-Rebbi mechanism still works (Fig. S2 in Supplementary Information). The wide choice of parameters provides more tunable condition for the acoustic realization.

Based on the tight-binding model discussed above, we now consider a PC implementation of the KM model with a Zeeman field along the $y$ direction for acoustic waves. The PC sample is fabricated by 3D printing, consisting of a bilayer structure with chiral cylinders for interlayer couplings. Each layer of the unit cell comprises two nonequivalent cylinders in a hexagonal lattice with a lattice constant $a = 3.55$ cm, connected by intralayer tubes (Fig. 2a). The radius and height of the cylinder in each layer are $r_0 = 0.4a/\sqrt{3}$ and $h_0 = 0.3a$. The side length of the square intralayer tube is $d = 0.18a$. The radius of the interlayer cylinder is $r_1 = 0.24a/\sqrt{3}$, and the interlayer distance is $h_1 = 0.4a$. Here, the interlayer chiral cylinder services two purposes: one is to supply the spin-orbit coupling, the other is to induce the Zeeman field. The bulk dispersion along the high symmetry lines is plotted in Fig. 2b. A complete band gap occurs around a frequency of 3.5 kHz.

The projected band dispersion on the zigzag boundary is demonstrated in Fig. 2c. The color maps denote the experimental dispersions, while the lines are for the simulated results. Gapped boundary state dispersions (solid white lines) appear in the projected bulk gap due to the in-plane Zeeman field. The experimental and simulated results agree well with each other. There are two kinds of configurations for the zigzag boundary corresponding to two nonequivalent cylinders on the outermost sites of the boundaries, respectively, shown as the left and right boundaries in Fig. 2d. Since the zigzag boundaries possess the $c_{2y}$ symmetry, the topological property of the boundary states can also be revealed by the quantized Zak phase. The simulated field distributions of the eigenmodes at the high symmetry point ($k_x = \pi/a$, marked by a red sphere in Fig. 2c) are shown in Fig. 2d. It is known that the eigenvalues of a two-fold rotation are $\pm 1$. That is to say, the eigenfield at $k_x = \pi/a$ remains unchanged or transforms with a $\pi$ phase under the $c_{2y}$ operation. The top panel of Fig. 2d displays one eigenmode, keeping invariant under the $c_{2y}$ operation. While in the bottom panel, the field localized on the opposite boundary, gains a $\pi$ phase under the $c_{2y}$ operation. These two modes correspond to two different eigenvalues of the $c_{2y}$ operation, resulting in the 0 and $\pi$ Zak phases of these two zigzag-boundary dispersions. This provides an alternative understanding of the topologically distinct zigzag boundaries, besides the Jackiw-Rebbi mechanism discussed in the lattice model.

The 1D topological boundary state dispersion guarantees the existence of the 0D corner states. Figure 3a shows the schematic diagram of a hexagon-shaped PC sample. Each corner is intersected by two boundaries with nonequivalent outmost cylinders. It is expected that the corner states should exist at all the six corners of the sample. The simulated eigenfrequencies of the hexagon-shaped sample are exhibited in Fig. 3b, where six localized states exist near 3.6 kHz in the gap of boundary state dispersions. The simulated corner states are also located at the right of the corners for the upper layer, consistent with the ones in the lattice model. The measured field in the left panel of Fig. 3c verified the existence of the corner state, which agrees well with the simulated result (right panel). Furthermore, we show the measured acoustic spectra of the corner with respect to frequency in Fig. 3d. The peak in the gap of boundary state dispersions (yellow region) around 3.6 kHz is due to the corner states.

We further explore the properties of the corner states in a hexagonal ring-shaped PC sample. Figure 4a shows the geometrical shape of the sample: six outer corners with 120° angle and six inner ones with 240° angle. Similar to the outer boundaries, the inner boundaries share the same topological properties, and result in corner states in the gap of the boundary state dispersions, as shown in Fig. 4b. However, the pressure field distributions are different. For the upper layer, the corner states at the outer boundary are identical to those in the above hexagon-shaped sample, i.e., locate at the right of the corners, while the corner states at the inner boundary locate at the left of the corners (yellow bars marked in Fig. 4a). The corner states for the lower layer locate at the opposite sides. More interestingly, the phase distributions of the outer corner remain unchanged under the $c_{2x}$ rotation, as shown by the simulated and measured pressure field distributions in Fig. 4c. But for the inner corner, the phase distributions are changed with a $\pi$ phase under the $c_{2x}$ rotation, as shown in Fig. 4d. These specific field distributions stem from the orthogonality of the corner states belonging to the outer and inner corners, since these two corners are complementary in geometry. These properties are consistent well with those in the lattice model, and the corner states are still stable even for a ring with a narrow width (Fig. S3 in Supplementary Information). In addition, the topological corner states also appear in a rhombus-shaped sample, and are robust against disorder (Fig. S4 in Supplementary Information).

In conclusion, we have realized a second-order TI phase with robust corner states, engineered from the acoustic KM model in the presence of a built-in Zeeman-like field along the $y$ direction. Our work confirms a new mechanism to realize the 2D second-order TI, which connects the $\mathbb{Z}_2$ TIs and second-order TIs, thus is of fundamental significance. Since the strength of the effective intrinsic spin-orbit coupling can be tuned at will by the chiral tubes, our system provides a useful platform to explore the underlying physics associated with the KM model. Moreover, the method of constructing the synthetic Zeeman field may pay a way to realize the Majorana-like corner states in acoustic systems[41,42].

**Acknowledgements**

This work is supported by the National Natural Science Foundation of China (Nos. 11890701, 11804101, 11704128, 11774275, 11974120, 11974005), the National Key R&D Program of China (Nos. 2018YFA0305800, 2017YFA0304203), the Guangdong Basic and Applied Basic Research Foundation (No. 2019B151502012), the Guangdong Innovative and Entrepreneurial Research Team Program (No. 2016ZT06C594).

**Author contributions**

All authors contributed extensively to the work presented in this paper.

**Competing financial interests**

The authors declare no competing financial interests.

**Methods**

**Numerical simulations.** All the numerical simulations of phononic crystals in Figs. 2-4 are calculated by using the commercial COMSOL Multiphysics solver package. The systems are filled with air, where mass density and speed of acoustic at room temperature are $\rho = 1.3 \text{ kg m}^{-3}$, and $v = 350 \text{ m s}^{-1}$, respectively. Because of the huge acoustic impedance contrast compared with air, the 3D printing materials are considered as rigid boundaries in the whole process of simulations. Periodic boundary conditions are applied in the $x$ and $y$ directions for calculating the bulk dispersions using one unit cell, and in the *y* direction for calculating the boundary state dispersion using ribbon structures. Rigid boundaries are used to calculate the corner states in the finite size samples.

**Experimental measurements.** The sample contains $30a \times 8a$ for the measurement of the boundary state dispersion. The hexagon-shaped sample has a length of side $22a$.

The lengths of side of the outer and inner boundaries of the hexagonal ring-shaped sample are $22a$ and $13a$, respectively. A sub-wavelength headphone is used to excite acoustic waves. The headphone is placed at the middle of the boundary of the sample for boundary state excitations, and it is placed at the corners of the hexagon-shaped and hexagonal ring-shaped samples for corner state excitations. A sub-wavelength microphone probe is inserted into the sample to measure the acoustic pressure fields. Acoustic signals are sent and recorded by a network analyser (Keysight 5061B). The pressure fields inside the samples are measured and Fourier transformed into the reciprocal space to obtain the boundary state dispersion.

**Data availability**

The data that support the plots within this paper and other findings of this study are available from the corresponding authors upon reasonable request.

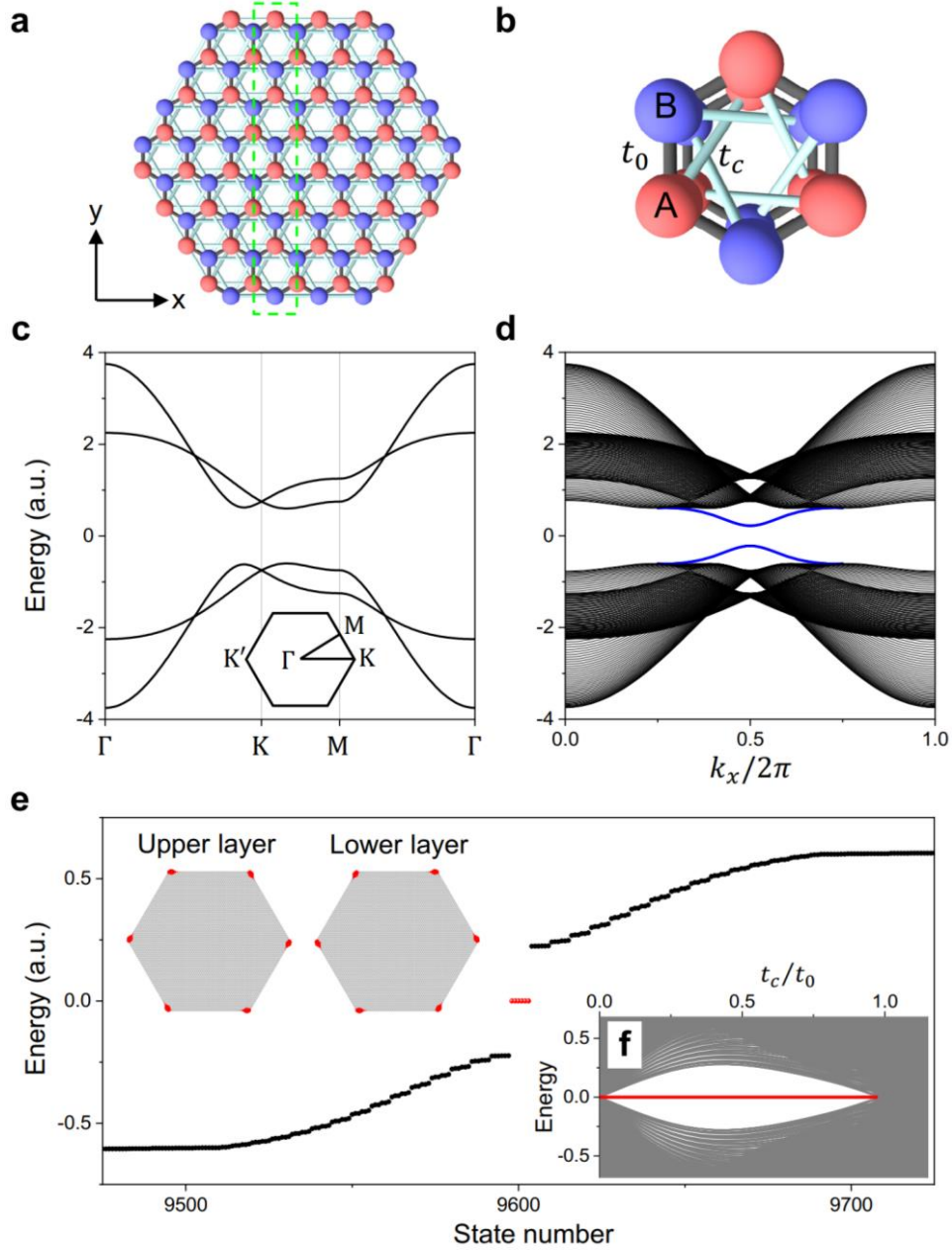

**Figure 1 | Corner states in the lattice model. a**, Schematic diagram of the hexagon-shaped sample of a bilayer hexagonal lattice. **b**, A unit cell with bilayer structure with sites A (red) and B (blue), the intralayer couplings $t_0$ (gray), and interlayer couplings $t_c$ (cyan). **c**, Bulk dispersion along the high symmetry lines. Inset: the first Brillouin zone. **d**, Projected dispersion of a ribbon in the presence of the zigzag boundaries denoted by the dashed box in **a**. **e**, Eigenvalues for the hexagon-shaped sample. The red spheres represent the corner states. Inset: Local density of states of the corner states for the upper and lower layers. The parameters above are chosen as $t_0 = -1$ and $t_c = t_0/4$. **f**, Eigenvalues of the hexagon-shaped sample as a function of $t_c/t_0$.

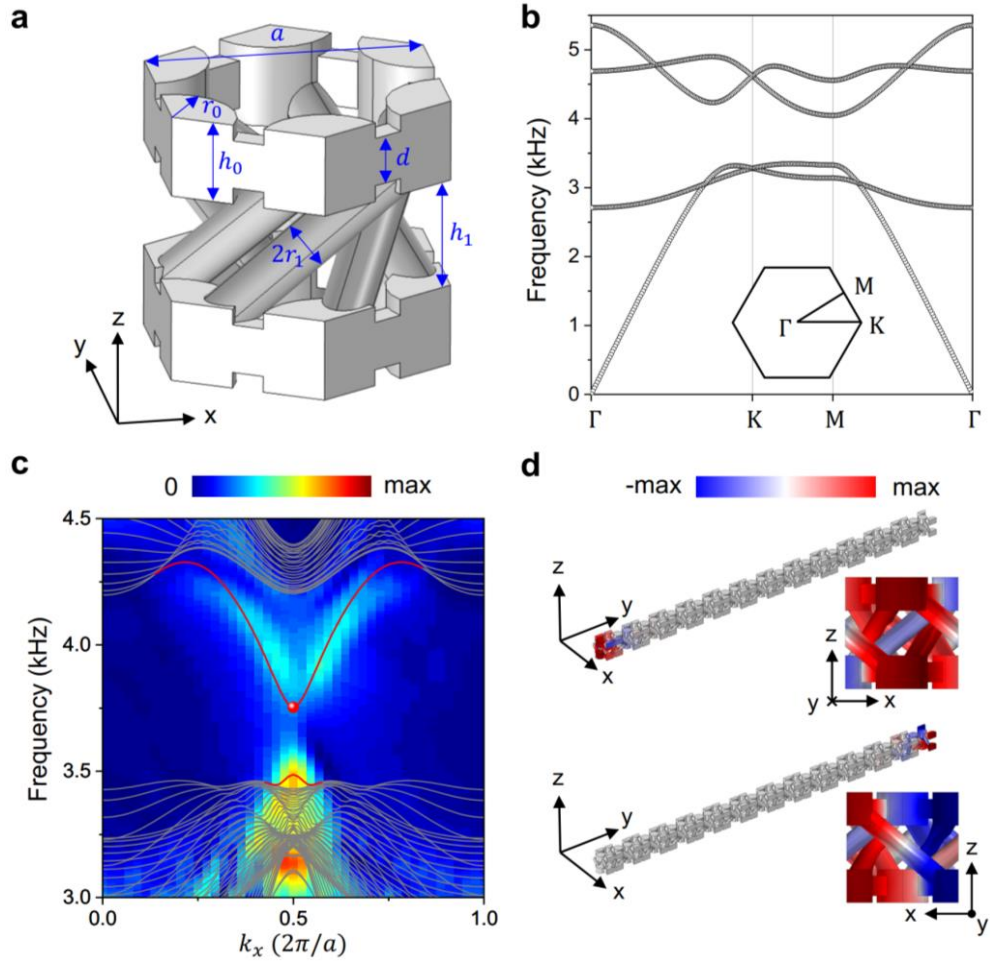

**Figure 2 | Bulk dispersion and the boundary state dispersions of the PC. a**, Schematic diagram of the unit cell. **b**, The bulk dispersions of the PC along the high symmetry lines. **c**, The simulated and measured boundary state dispersions, represented by the solid lines and color maps, respectively. For the simulated results, the black lines denote the projected bulk dispersions and the white lines are for the boundary state dispersions. **d**, The pressure field distributions of the eigenmodes at $k_x = \pi/a$ (marked as a red sphere in **c**), for two kinds of zigzag boundaries.

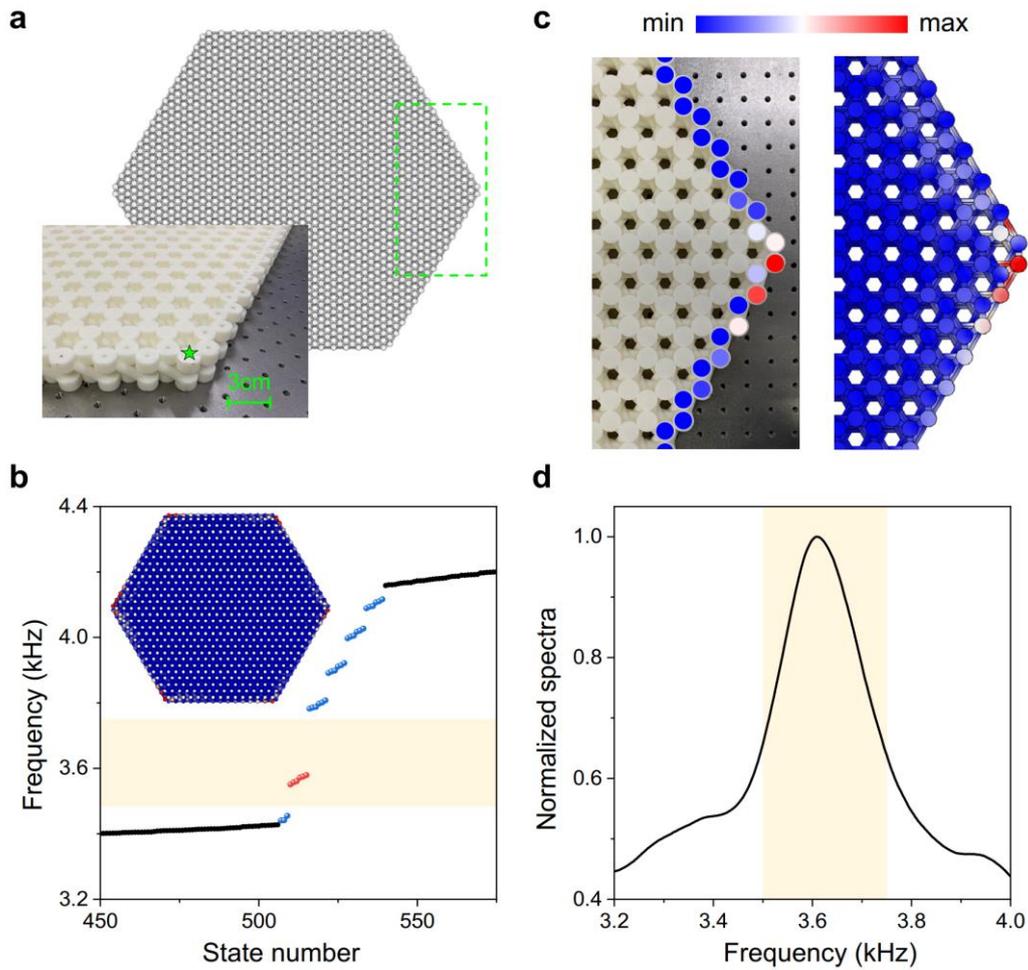

**Figure 3 | Corner states in a hexagon-shaped PC. a**, The schematic diagram for a hexagon-shaped PC. Inset: Partial view of the PC sample. The green star denotes the position of exciting source. **b**, Simulated eigenfrequencies of the PC. The red, blue and black spheres represent the corner, edge and bulk states, respectively. Inset: Simulated pressure field distribution of one corner mode. **c**, The measured (left panel) and simulated (right panel) pressure field distributions of the corner marked by the dashed rectangle in **a** at 3.6 kHz. **d**, The measured acoustic spectra as a function of frequency.

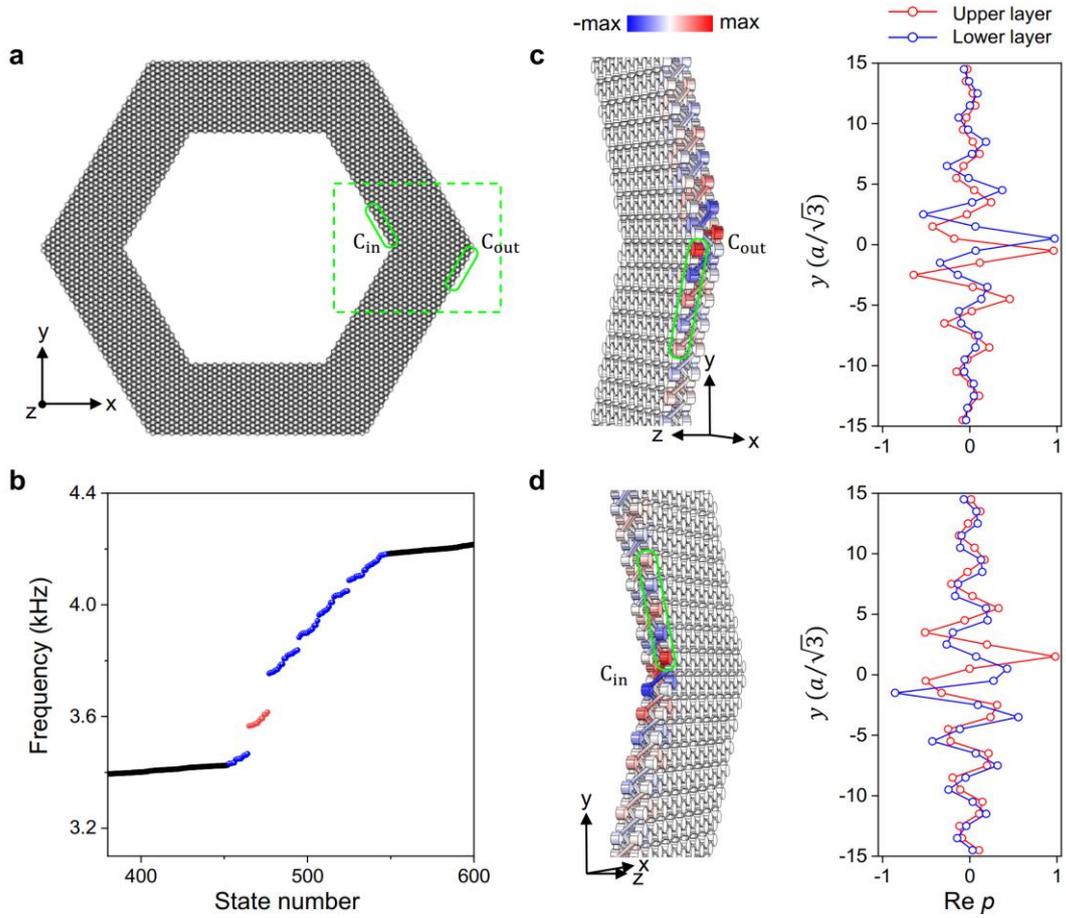

**Figure 4 | Corner states at the outer and inner boundaries of a hexagonal ring-shaped PC. a**, The schematic diagram for a hexagonal ring-shaped PC sample. A pair of inner and outer corners are denoted by $C_{in}$ and $C_{out}$. The yellow bars label the positions of maximum amplitudes of corner states for the upper layer. **b**, Simulated eigenfrequencies of the PC. There are twelve corner modes (red) in the gap of edge states (blue). **c**, The simulated (left panel) and measured (right panel) pressure field distributions in the $C_{out}$ corner boxed by the dashed rectangle in **a** at the frequency of 3.6 kHz. **d**, The corresponding results for the $C_{in}$ corner.